# A non-invasive investigation of Egyptian faience using a long wavelength optical coherence tomography (OCT) at 2μm

Margaret Read[a, b], Chi Shing Cheung[a], Haida Liang[a,*], Andrew Meek[b], Capucine Korenberg[b]

[a] School of Science and Technology, Nottingham Trent University, Nottingham, NG11 8NS

[b] Department of Scientific Research, The British Museum, London, WC1B 3DG

*Corresponding author: Haida.Liang@ntu.ac.uk

*Egyptian faience is a non-clay ceramic semi-transparent material, formed of a quartz core and alkali-lime glaze with some cases exhibiting an interaction layer between them. Several possible glazing methods have been identified. Previous investigations have tried to identify the glazing technique by using the microstructure images obtained from polished sections using scanning electron microscope (SEM). Such techniques require sampling which is not feasible on museum collections. Optical Coherence Tomography (OCT) is a non-invasive 3D imaging technique, that produces virtual cross-sections of transparent and semi-transparent materials. Liang et al. (2012a) investigated the feasibility of using OCT to non-invasively investigate microstructures of Egyptian faience, but the limited probing depth of the 930nm OCT prevented viewing down to the core of the objects, where the presence of glass was thought to be a distinguishing feature between some of the manufacturing techniques, and where the particle size of the quartz may indicate the difference in the raw material. In this paper, a unique longer wavelength OCT at 2μm is used to scan a number of ancient Egyptian faience objects including ring and shabti fragments. It was found that the core of the faience could be imaged at this longer wavelength, allowing comparisons in all the layers within the microstructure, and leading to discussions about the possible glazing methods. The 2μm OCT offers the possibilities of rapid, non-invasive imaging of faience microstructure down to the core, allowing comprehensive studies of intact objects and large museum collections.*

## 1. Introduction

Ancient Egyptian faience is a non-clay ceramic material consisting of a fine- or coarse-grained silica (quartz) core, an alkali-lime glaze and, in most cases, an interaction layer connecting these layers. The glazing methods for faience have been categorised into three main types: efflorescence, cementation and application (Nicholson and Peltenberg, 2000). These types are defined using the microstructure of the faience through scanning electron microscope (SEM) images of polished sections extracted from the fragments (Tite et al. 1983). The cementation technique (Fig.1a,d) buries a formed quartz body in the glazing salts. During the firing process the glaze forms on the outside of the object and therefore there is unlikely to be inter-particle glass in the core (Tite et al. 1983). The efflorescence technique (Fig.1b,e) mixes the glazing salts into the quartz body which is moulded and then fired; as the glazing salts are mixed within the core, this method results in vitrified material, known as inter-particle glass, within the core after firing. The application technique (Fig.1c) applies the glaze onto the quartz body as a slurry; the glaze slurry is applied to the surface; therefore, inter-particle glass is not found within the core after firing. The SEM images of the polished sections of laboratory replicas made with the 3 different techniques clearly shows that the glazing techniques and the particle size of the raw material can significantly influence the microstructure. Tite *et al.* (1983) suggested that the objects can be differentiated and categorised by the presence of glass within the core. However, there has been much debate and speculation about how valid this method is. In particular, the glazing methods could have been used in combination with each other (Vandiver 1998, Vandiver 1983, Nicholson 1993, Tite et al. 1983), for example, mixing interparticle glass into the core to reinforce the structures of delicate objects, such as rings. This means the glazing method would have more to do with the size or purpose of the object than a time evolution or regional difference in technique. Nevertheless, the non-existence of interparticle glass in the core can exclude efflorescence as a possible glazing method. While it may not be possible to determine with certainty the glazing method given the potential complications of hybrid techniques, microstructure still gives valuable information allowing at least the sorting of objects into groups according their microstructure. Those with similar microstructure is most likely to have used similar raw material and undergone similar glazing method.

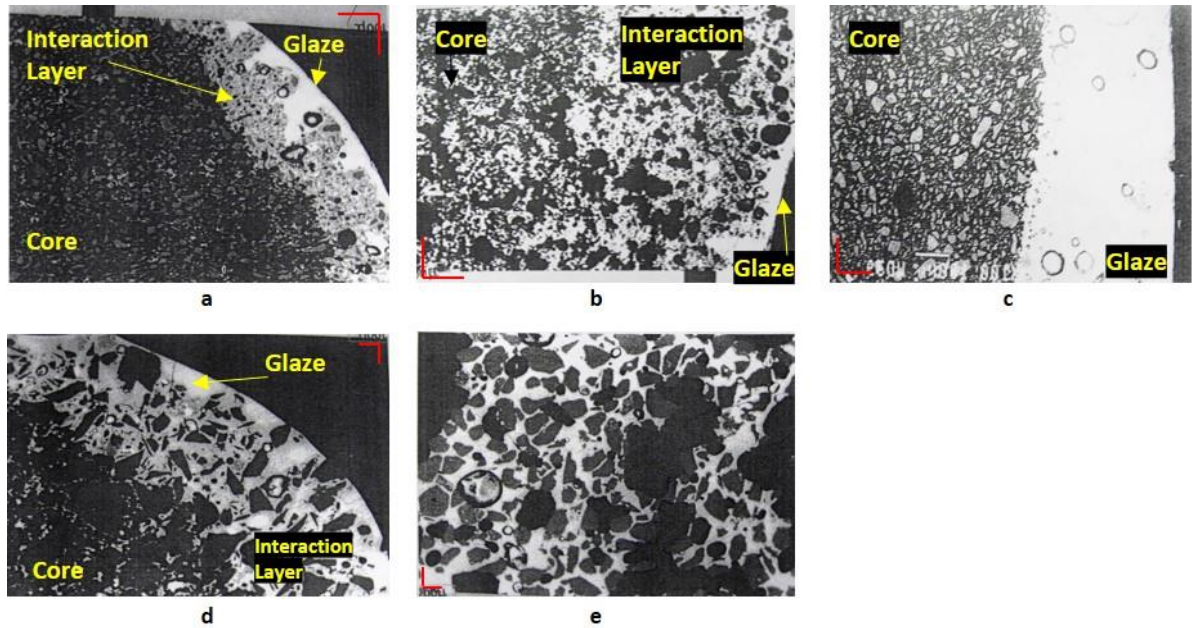

*Figure 1.* SEM images of polished sections of lab made replicas using various glazing methods. White areas are glaze, grey areas are quartz and black areas are pores; red scales equal 100μm; (a) Cementation Method – Fine-grained quartz (Tite et al., 1983). (b) Efflorescence Method – Fine-grained quartz (Tite et al., 1983). (c) Application Method - Fine-grained quartz (Tite and Bimson, 1986). (d) Cementation Method – Coarse-grained quartz (Tite et al., 1983). (e) Efflorescence Method - Coarse-grained quartz core (Tite et al., 1983).

However, the method employing SEM (Tite et al. 1983) requires sampling, which is an issue for museum objects, preventing the study of large collections of intact objects. Previous investigations (Liang *et al.,* 2012a; Liang *et al.,* 2012b) demonstrated the feasibility of using OCT, a non-invasive 3D volume imaging technique, to image the subsurface microstructure of Egyptian faience without sampling. However, the OCT operating at 930nm, which happens to be within the absorption band of $Cu^{2+}$ ions that gives the typical turquoise colour of the faience objects, had limited probing depth that allowed only the glaze and interaction layers to be imaged in most cases. A long wavelength OCT at 2μm has since been developed to improve the probing depth by moving away from the absorption bands of most of the colorants and moving to a longer wavelength where optical scattering is reduced (Cheung et al. 2015). The use of this OCT system to compare and contrast the microstructure of ancient Egyptian faience artefacts is investigated in the present study.

## 2. Materials and Methods

### 2.1 Ancient Egyptian faience objects

The Egyptian faience objects used in this investigation were taken from a collection of reference objects, dating from approximately 1600 – 300BC, housed in the British Museum's Department of Scientific Research. The collection contained broken fragments of rings and shabtis of a mixture of ages, all turquoise/blue in colour. Five fragments of Egyptian faience (Table 1) were selected and imaged with multiple areas of 6mm x 7mm on each object.

| BMRL No. | Object Type | Date |
| --- | --- | --- |
| 4698-16320-M | Ring | New Kingdom (c. 1550-1077 BCE) |
| 4698-16321-P | Ring | 18th Dynasty (c. 1550-1292 BCE) |
| 4698-16322-R | Shabti | Late Period (c. 664-332 BCE) |
| 4698-16323-W | Shabti | 21st Dynasty (c. 1069-945 BCE) |
| 4698-16324-Y | Shabti | 21st Dynasty (c. 1069-945 BCE) |

Table 1: Five items surveyed from the British Museum Department of Scientific Research Reference Collection

An earlier study by Tite *et al.* (1983) used SEM on polished sections from the same reference objects and replica beads (Tite et al. 1983; Tite and Bimson 1986), but the requirement to destructively sample the objects for SEM analysis limits the scope of the objects available to survey and each polished section only provides a single cross-section. Previous investigations by Liang *et al. (*2012a*,* 2012b) used the same British Museum Reference Collection objects and faience beads produced in the laboratory to demonstrate the feasibility of using 930nm OCT to study the glaze of Egyptian faience without sampling. However, limited probing depth prevented the OCT from imaging all the layers of the Egyptian faience microstructure.

### 2.2 Long wavelength OCT at 2um

OCT is based on a Michelson interferometer with a broadband laser source. The OCT used in this study was a 2μm Fourier Domain OCT (Cheung et al., 2015), consisting of a supercontinuum laser source with a bandwidth of 220nm resulting in a depth resolution of ~7μm in material (assuming the refractive index is ~1.5), a fixed reference path and a high resolution spectrometer. The transverse resolution given by the objective lens is 17μm. The interference signal essentially compares the path travelled by the photon scattered back from the object with the reference path to measure the depth of the layers. This interference signal is recorded as a spectrum which is then Fourier transformed to create a depth profile. Scanning of the laser beam over a line segment produces a virtual cross-section image and scanning over an area produces the 3D microstructure of the surface and subsurface volume in the form of a 3D image cube. The speed of capturing a typical image cube of 500 by 500 depth profiles is ~2 minutes.

OCT measures optical distances, or the time it takes for the light to travel through the medium, in the depth direction (or direction of the optical axis of the OCT). Physical depth or thickness, is measured by dividing the optical thickness by the refractive index. All the OCT images presented in this paper are 1.3 mm in optical depth. A number of areas, each of 6mm x 7mm, were scanned per object. The OCT was focussed at ~500 μm from the top of the image.

OCT is sensitive to changes in refractive index (RI), as larger changes in refractive index means higher reflectivity as governed by Fresnel equations. The brighter areas in an OCT image represents regions of high reflectivity, and dark areas are where there is little change in refractive index and therefore very few photons are scattered back from these regions.

## 3. Results and Discussion

Using the 2μm OCT allowed new aspects of the microstructure of Egyptian faience to be investigated. This is mostly due to the increased probing depth of imaging in the 2μm regime, which can view down to the quartz core. The increased probing depth coupled with the use of an OCT image cube can show characteristics of each layer throughout the microstructure of the object. The important aspect that the 2μm OCT shows is how very different the microstructures of the objects appear in the virtual cross-section images.

Even at first glance the OCT images of the Egyptian Faience show differences in microstructure, however, the interpretation of an OCT image is key to the understanding of exactly how the microstructures differ from object to object. As mentioned, OCT is sensitive to the change in refractive index (RI) between media, as well as this, the orientation and particle size of the media can affect the appearance of the OCT image. Simplistically, in terms of RI, the Egyptian Faience has three components: glass (glaze), silica (quartz), air bubbles in the otherwise homogeneous glaze or pores in the core or interaction layer. At 2μm, the RI of soda-lime glass is ~1.50 (Rubin 1985), fused silica quartz is ~1.44 (Malitson 1965) and air is ~1. When the RI change is large, for example, when the light reaches the air to glaze (or glaze/air bubble) interface, a bright well defined line appears. If the difference in RI is small, for example, between quartz and glaze, the OCT will show varying levels of weak scattering, depending on the grain size of the quartz and the orientation of the surface of a grain relative to the optical axis. When there is no change in RI, for example, within the glaze of the Egyptian Faience, or within a large grain of quartz or large pore, the area in the OCT image will be dark.

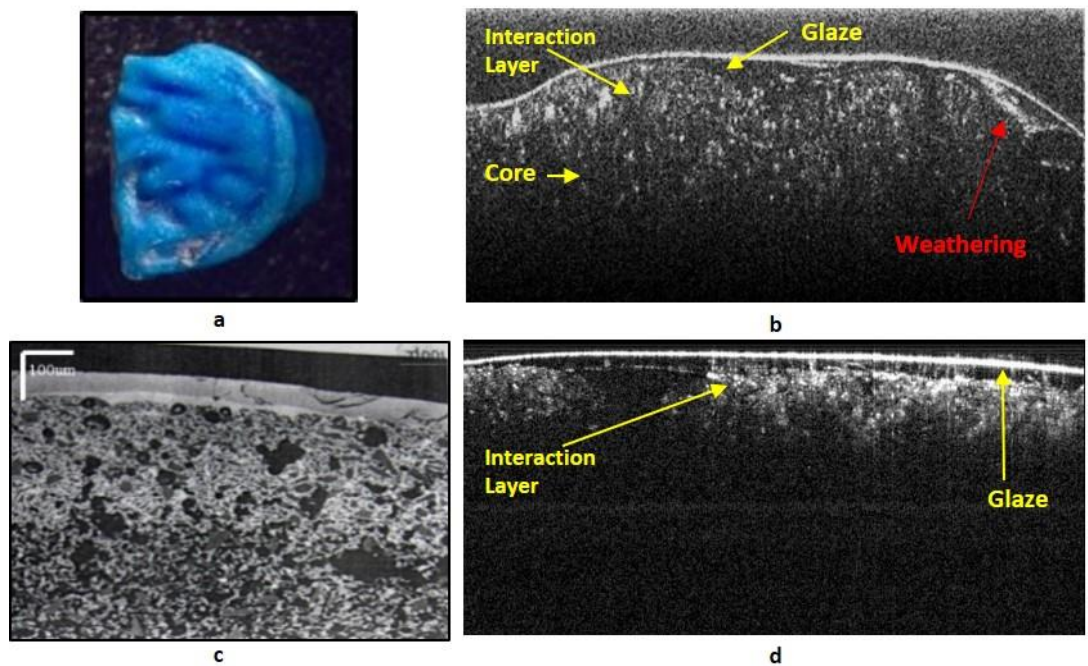

*Figure 2.* A 18[th] Dynasty ring (BMRL 4698-16321-P) from the British Museum Department of Scientific Research Reference Collection (a) Photograph of the fragment, copyright is to the British Museum Department of Scientific Research (b) 2μm OCT virtual cross-section of the ring fragment and a video of an area scan; (c) SEM image of polished section of the ring fragment (Tite et al., 1983). Scales equal 100μm (d) 930nm OCT image (Liang et al., 2012). Dimensions of the OCT images are 1.3mm x 6mm (depth x width).

The advantage the 2μm OCT has over the 930nm OCT is that all three layers can be imaged, rather than, in most cases, 2 layers (Figures 2, 3 and 4). However, this increased probing depth does have a trade-off, the depth resolution has decreased from ~4.5μm (930nm OCT) (Liang *et al.*, 2012a) to ~7μm (2μm OCT) (Cheung, 2015), which means any small features or details close to each other such as thin gel layers (<7μm) in the glaze due to weathering may become unresolvable. On the other hand, the difference in depth resolution between the two OCT systems have not made much difference to the visibility of any of the small features in the faience examples given here. Another trade-off is the image contrast, where it is lower at 2μm than 930nm, since scattering is lower at longer wavelength, which is also why the depth of penetration is in general greater at 2μm.

The 2μm OCT image of the 18$^{th}$ Dynasty ring (BMRL 4698-16321-P) (Fig. 2b) shows, at the top of the image, the air-glass interface and a well-defined transparent glaze layer. The bright dots in the interaction layer, can be interpreted as very small air bubbles in the glass matrix which can also be seen directly in the SEM image. The low-level scattering background in this layer most likely corresponds to the scattering at interfaces of the quartz particles in the glass matrix. Judging by the interaction layer in the OCT image it has fine-grained quartz particle compared with the faience shown in Fig. 3, which agrees with what is seen in the SEM images (Fig. 2c and 3c). The core shows a decreased level of scattering compared with the interaction layer suggesting reduced number of air bubbles, pores and possibly the presence of interparticle glass.

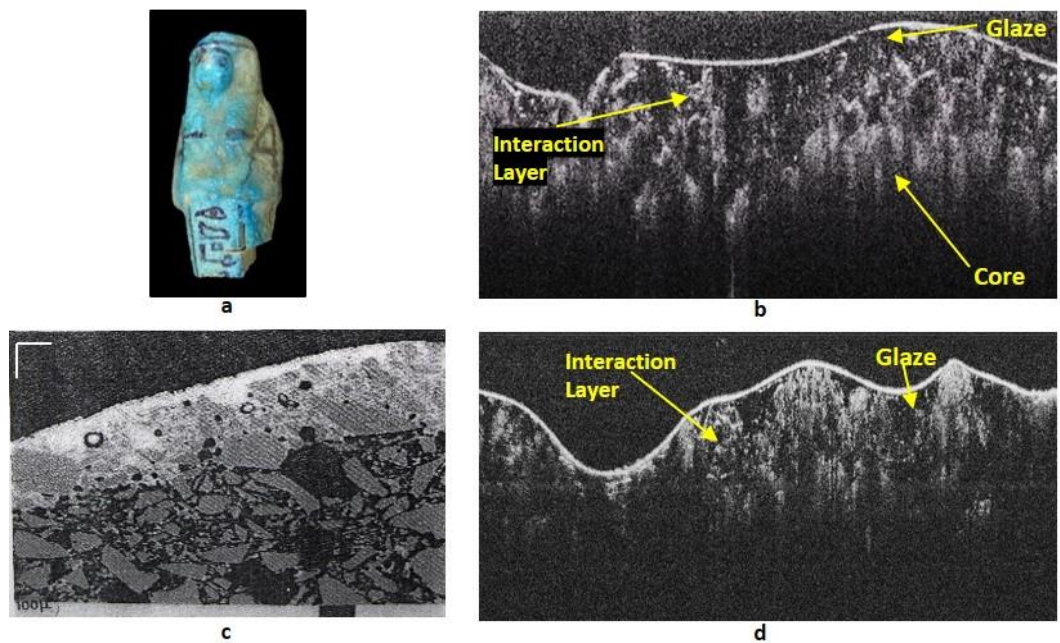

***Figure 3.*** *A 21$^{st}$ Dynasty shabti (BMRL 4698-16323-W) from the British Museum Department of Scientific Research Reference Collection (a) Photograph of the shabti, , copyright is to the British Museum Department of Scientific Research (b) 2μm OCT virtual cross-section image of the shabti and a video of an area scan;(c) SEM image of polished section of the shabti (Tite et al., 1983) scales equal 100μm (d) 930nm OCT image (Liang et al., 2012). Dimensions of the OCT images are 1.3mm x 6mm (depth x width).*

The images of the 21st Dynasty shabti (BMRL 4698-16323-W) (Fig.3) shows a very different structure to that seen in the 18th Dynasty ring (Fig.2). At the top of the image, there is a bright air-glass interface followed by pockets of glaze of varying thickness. There are air bubbles and large quartz particles present in the interaction layer. The stronger scattering in the core is an indication of more air/quartz interfaces in the core and a less homogeneous core compared with the ring in Fig. 2, which is consistent with the SEM images. This object has coarse grained quartz judging by the structure of the interaction layer. The scattering in the core is so strong that it is dominated by multiple scattering preventing the quartz particles to be seen directly. The OCT image is consistent with the SEM image: the irregular glaze thickness, air bubbles in the interaction layer and coarse-grained quartz particles.

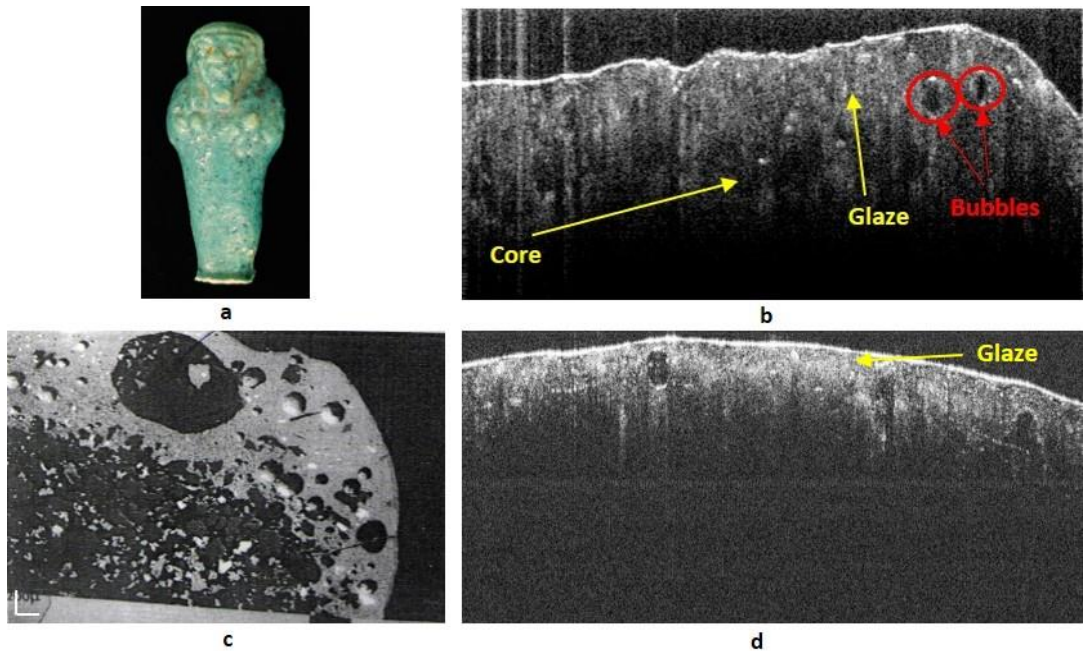

*Figure 4.* A Late Period shabti (BMRL 4698-16322-R) from the British Museum Department of Scientific Research Reference Collection (a) Photograph of the shabti, copyright is to the British Museum Department of Scientific Research (b) 2µm OCT virtual cross-section image of the shabti and a video of an area scan; (c) SEM image of polished section of the shabti (Tite et al., 1983), scales equal 100µm; (d) 930nm OCT image (Liang et al., 2012). Dimensions of the OCT images are 1.3mm x 6mm (depth x width).

The OCT image of the late period shabti fragment (BMRL 4698-16322-R) (fig. 4b) again shows a very different structure to what is seen in the previous two objects (fig. 2 & fig.3). The top of the image shows the bright air/glass interface, under which there is a very high number of bright dots indicating air bubbles of varying sizes. Unlike the earlier examples, the glaze is not a transparent layer but rather a uniformly scattering layer which may be due to the small white high atomic number particles seen in the thick glaze layer in the SEM image. There is no interaction layer in the OCT image, but a dark band is seen at the beginning of the core layer which is consistent with the SEM image.

Images of another shabti (BMRL 4698-16324-Y) (Fig. 5a) and ring (BMRL 4698-16320-M) (Fig. 5b) show similarities with the other objects discussed above. The OCT images of the 21st Dynasty shabti (Fig. 5a) shows very close similarities to that seen in the first 21st Dynasty shabti (Fig. 3). The OCT images of the new Kingdom ring (Fig. 5b) show very close similarities to that seen in the 18th Dynasty ring (Fig. 2).

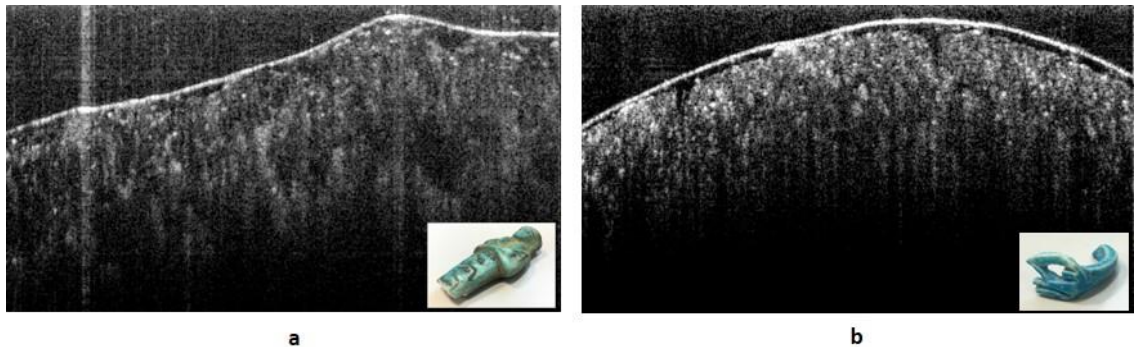

*Figure 5.* 2µm OCT virtual cross-section images of two objects from the British Museum Department of Scientific Research Reference Collection that were also scanned; dimensions are 1.3mm x 6mm (depth x width). The copyright of the Egyptian faience photographs is to the British Museum Department of Scientific Research (a) 21st Dynasty shabti (BMRL 4698-16324-Y). (b) New Kingdom ring fragment (BMRL 4698-16320-M).

In summary, OCT images can be used to sort the faience microstructures into 3 groups. The OCT images show very different microstructures between the 18th Dynasty ring (Fig. 2), 21st Dynasty shabti (Fig. 3) and the Late Period shabti (Fig. 4). The images also show similarities between the 21st Dynasty shabtis (Fig.3 & Fig. 5a) and between the 18th Dynasty and New Kingdom rings (Fig.2 & Fig.5b). Even though OCT has a lower resolution compared with SEM and the interpretations of OCT images may not be as straightforward, the advantages of OCT being a rapid, non-contact and non-invasive technique, means in the case of Egyptian faience, OCT is much more useful than SEM. OCT allows intact objects to be imaged over large enough area to be representative of the whole object, while SEM needs sampling and each polished section gives only one cross-section image.

**4. Conclusion**

A longer wavelength OCT provides an increased probing depth and could image all the layers of Egyptian faience samples examined in this investigation. This fact combined with the non-contact and non-invasive nature of the technique, rapid imaging speed that allows multiple areas of acquisition, makes the 2µm OCT an advantageous technique for studying the microstructure of Egyptian faience.

Many of the limitations in the identification of the manufacturing techniques of the Egyptian faience based on microstructure are innate to the objects which are more complicated to categorise due to hybrid glazing techniques. The advantage of using long wavelength OCT is to be able to compare all 3 layers (glaze, interaction and core) of the microstructure in a larger selection of faience examples. This allows discussion on comparisons between particle size, glaze thickness, indications of the presence of interparticle glass and whether there is an interaction layer. The 2µm OCT allows the grouping of faience objects based on the full information (all 3 layers) on their microstructure. Such groupings will reflect the similarity of raw material and manufacture technique and allow the studies of the correlation between microstructure and object type or size; correlations between microstructure and geographic location and/or historic period.

**Acknowledgements**

Funding from the UK Arts and Humanities Research Council (AHRC) Collaborative Doctoral Programme is gratefully received. The development of the 2μm OCT was funded by the UK AHRC and Engineering and Physical Sciences Research Council (EPSRC) Science & Heritage Programme (Interdisciplinary Research Grant AH/H032665/1). We are grateful to NKT Photonics for loan of the supercontinuum laser when our own laser was being repaired.